\pdfoutput=1
\documentclass[ journal,article,accept,moreauthors,pdftex]{Definitions/mdpi} 

\usepackage{subcaption}
\usepackage{color}
\firstpage{1} 
\pubvolume{1}
\issuenum{1}
\articlenumber{0}
\pubyear{2021}
\copyrightyear{2020}
\datereceived{} 
\dateaccepted{} 
\datepublished{} 
\hreflink{https://doi.org/} 

\Title{Exploiting Pre-Trained ASR Models for Alzheimer's Disease Recognition Through Spontaneous Speech}

\TitleCitation{Title}

\Author{Ying Qin$^{1,\ddagger}$, Wei Liu $^{2,\ddagger}$, Zhiyuan Peng  $^{2,\ddagger}$, Si-Ioi Ng$^{2}$, Jingyu Li$^{2}$, Haibo Hu$^{1}$ and Tan Lee $^{2*}$}

\AuthorNames{Firstname Lastname, Firstname Lastname and Firstname Lastname}

\AuthorCitation{Lastname, F.; Lastname, F.; Lastname, F.}

\address{%
$^{1}$ \quad Institute of Information Science, Beijing Jiaotong University, Beijing 100044, China; yingqin@bjtu.edu.cn\\
$^{2}$ \quad Department of Electronic Engineering, The Chinese University of Hong Kong, Hong Kong, China; \{louislau\_1129, siioing, lijingyu0125\}@link.cuhk.edu.hk, jerrypeng1937@gmail.com
}

\corres{Correspondence: tanlee@cuhk.edu.hk}

\secondnote{These authors contributed equally to this work.}

\abstract{Alzheimer's disease (AD) is a progressive neurodegenerative disease and recently attracts extensive attention worldwide. 
Speech technology is considered a promising solution for the early diagnosis of AD and has been enthusiastically studied. 
Most recent works concentrate on the use of advanced BERT-like classifiers for AD detection. Input to these classifiers are speech transcripts produced by automatic speech recognition (ASR) models. The major challenge is that the quality of transcription could degrade significantly under complex acoustic conditions in the real world. The detection performance, in consequence, is largely limited. This paper tackles the problem via tailoring and adapting pre-trained neural-network based ASR model for the downstream AD recognition task.
Only bottom layers of the ASR model are retained. A simple fully-connected neural network is added on top of the tailored ASR model for classification. The heavy BERT classifier is discarded. The resulting model is light-weight and can be fine-tuned in an end-to-end manner for AD recognition. Our proposed approach takes only raw speech as input, and no extra transcription process is required. The linguistic information of speech is implicitly encoded in the tailored ASR model and contributes to boosting the performance. Experiments show that our proposed approach outperforms 
the best manual transcript-based RoBERTa by an absolute margin of $4.6\%$ in terms of accuracy. Our best-performing models achieve the accuracy of $83.2\%$ and $78.0\%$ in the long-audio and short-audio competition tracks of the 2021 NCMMSC Alzheimer's Disease Recognition Challenge, respectively.}


\begin{document}
\section{Introduction}
Alzheimer’s disease (AD) is a neurodegenerative disease with a progressive pattern of cognitive and functional impairment. 
Symptoms like language impairment, memory loss, self-neglect, and behavior issues are found in patients as the disease worsens. 
A tremendous amount of effort has been made toward early detection of the disease \cite{ritchie2017midlife}.
Studies in \cite{yu2015cognitive,ivanov2013phonetic,luz2018method,haider2019assessment,luz2020alzheimer,mirheidari2018detecting} 
suggested the early symptoms of AD were detectable from spontaneous speech. 
Technologies that enable automatic analysis of human speech are considered a promising solution to this problem.




Existing methods of automatic AD detection 
can be divided into audio-based \cite{haider2019assessment,luz2018method,ivanov2013phonetic,yu2015cognitive,ambrosini2019automatic} and transcript-based \cite{mirheidari2018detecting,li2021comparative,ye2021development,Syed2020} approaches.
Conventional audio-based methods exploit acoustic, articulatory, phonetic and prosodic features in speech signal. 
Impairment in speech content is neglected in the modeling process.
Studies in \cite{li2021comparative,Luz2021a,DBLP:conf/interspeech/BalagopalanERN20} suggest such audio-based method shows less  satisfactory performance than the transcript-based method. 
The transcript-based methods adopt various type of features derived from text content to characterize language impairment. Example features are part-of-speech, vocabulary richness and syntactic complexity features \cite{2015Linguistic}. 
It requires laborious efforts to manually transcribe speech into text.
Redundant information in the raw recording, such as background noise, channel distortion, gender and voice variation, etc., is removed in the transcription process.
Evidences \cite{yuan2020disfluencies,Luz2021a} also show that, with additional annotations of pause and speech dis-fluency, the performance could be further improved.
However, the preparation for manual transcripts is time-consuming and costly, a fully automatic pipeline for AD detection is highly desirable.
Replacing human transcribers by automatic speech recognition (ASR) systems is naturally expected to be a promising solution\cite{Rohanian2021,ye2021development}. 
The major challenge is that the quality of transcription may degrade significantly on speech with atypical accent and acoustic conditions, not to mention the speech coming from language impaired patients.
This urges more efficacious way of integrating ASR models into the analysis and detection processes.

Studies on this problem  \cite{ye2021development,Rohanian2021,Gauder2021,Pan2021} have been largely focused on taking advantage of the superior back-end classifier, e.g., the BERT model \cite{devlin2018bert}, to analyze the transcripts given by ASR. 
Adaptation of the front-end ASR model is under-investigated. 
As a matter of fact, 
the back-end system performance is sensitive to the quality of transcripts generated by the front-end ASR system.
If the 
ASR system encounters more challenging conditions, e.g. the test speech has different accents and dialects, or is masked by background noise, the performance of the back-end system is further limited.

The present study aims to address this problem by adaptation of pre-trained neural-network based ASR models. 
Bottoms layers of the front-end ASR model  \cite{baevski2020wav2vec,zhang2020transformer} are retained. The original heavy back-end classifier like BERT is discarded and replaced by a simple fully-connected neural network. 
The front-end and the back-end are integrated into a single light-weight model, which can be optimized jointly for end-to-end AD recognition. 


In this paper, two representative end-to-end (E2E) ASR models, namely the supervised joint CTC-attention model\cite{zhang2020transformer} and the self-supervised wav2vec2.0 based model \cite{baevski2020wav2vec}, are investigated. 
Extensive experiments are conducted on the dataset provided by the 2021 NCMMSC Alzheimer's Disease Recognition Challenge. The proposed method circumvents the use of manual transcripts, and outperforms the best transcript-based method with RoBERTa \cite{liu2019roberta} by an absolute improvement of $\mathbf{4.6}\%$ in terms of accuracy.

The rest of this paper is organized as follows. Section \ref{sec:data} describes the data. Section \ref{sec:approach} explains the baseline systems and our proposed system design. Experimental setup and results are given in Section \ref{sec:exp-setup} and Section \ref{sec:exp-result}, respectively. Section \ref{sec:conclusion} concludes this paper.

\section{Data Description}
\label{sec:data}
\subsection{The dataset and data clean-up}
The dataset is provided by the \textbf{2021 NCMMSC Alzheimer’s Disease Recognition Challenge} (referred to as ``AD dataset''). 
It contains audio samples of natural speech from a total of 124 speakers, which are divided into three groups: Alzheimer's disease (AD, $26$ speakers with $79$ samples), mild cognitive impairment (MCI, $54$ speakers with $93$ samples) and healthy control (HC, $44$ speakers with $108$ samples). The length of each audio sample is about $30$-$60$ seconds. 
The speech content includes picture description, fluency test and free conversation. 
Most subjects speak in different Chinese accents and dialects. 

Data inspection and cleaning were carried out. A few duplicated audio files of one speaker (AD\_F\_040108) were identified and removed (recording ID: $045, 046$ and $047$). Another audio sample (HC\_M\_019216\_001) which sounds abnormal were also removed.  

\subsection{Data partition}
After data cleaning, the audio samples in the dataset are partitioned into three subsets, namely training set (\textit{train}), development set (\textit{dev}) and test set (\textit{test}) according to the split ratio of $0.70:0.15:0.15$. 
In addition, we notice strong data imbalance in the AD group, i.e., $47$ out of $79$  speech samples are from the speaker AD\_F\_040108.

The strategy of manual split was adopted to better control the data balance across the subsets at various levels, such as  gender (male/female) and group (AD/MCI/HC). 
Note that the speakers in the \textit{test} set are totally unseen to the \textit{train} and  \textit{dev} sets. 
Partial overlap of speaker identities 
between the \textit{train} set and  \textit{dev} set is allowed. The speaker diversity was enlarged as far as possible in the \textit{train} and \textit{test} sets.

To examine the effect of various data-split results on model development,
we re-distribute the number of speakers and samples between the  \textit{train} set and the \textit{dev} set twice, while keeping the  \textit{test} set unchanged. As a result, we obtain three versions of manual splits named v1,v2 and v3, respectively. The detailed information about data partition is shown in  Table \ref{dataset}. 
\begin{specialtable}[t] 
\caption{The statistical information of the splitted AD dataset, the number after \# indicates the amount of speech audio samples and the number in bracket denotes the amount of speakers. The prefix $\ast$ represents the \textit{dev} set has partial speaker overlap with the \textit{train} set. \label{dataset} }
\scalebox{0.8}{
\begin{tabular}{c|ccc|ccc|ccc}
\toprule
\multirow{2}{*}{splits} & \multicolumn{3}{c|}{\textit{train}}        & \multicolumn{3}{c|}{$\ast$ \textit{dev} }         & \multicolumn{3}{c}{\textit{test}}          \\ \cline{2-10}
                        & AD        & MCI       & HC        & AD       & MCI       & HC        & AD        & MCI       & HC        \\ \midrule \midrule
\multirow{2}{*}{v1}     & \#52 (15) & \#65 (38) & \#74 (29) & \#12 (3) & \#14 (10) & \#16 (10) & \#12 (10) & \#14 (10) & \#17 (11) \\ \cline{2-10} 
                        & \multicolumn{3}{c|}{\#191 (82)}   & \multicolumn{3}{c|}{\#42 (23)}   & \multicolumn{3}{c}{\#43 (31)}     \\ \hline
\multirow{2}{*}{v2}     & \#52 (11) & \#60 (34) & \#71 (24) & \#12 (9) & \#19 (9)  & \#19 (8)  & \#12 (10) & \#14 (10) & \#17 (11) \\ \cline{2-10} 
                        & \multicolumn{3}{c|}{\#183 (69)}   & \multicolumn{3}{c|}{\#50 (26)}   & \multicolumn{3}{c}{\#43 (31)}     \\ \hline
\multirow{2}{*}{v3}     & \#55 (16) & \#69 (40) & \#80 (30) & \#9 (3)  & \#10 (7)  & \#10 (6)  & \#12 (10) & \#14 (10) & \#17 (11) \\ \cline{2-10} 
                         & \multicolumn{3}{c|}{\#204 (86)}   & \multicolumn{3}{c|}{\#29 (16)}   & \multicolumn{3}{c}{\#43 (31)}     \\ \bottomrule
\end{tabular}
}
\end{specialtable}

\section{Approaches}
\label{sec:approach}
\subsection{Baselines}
\subsubsection{Conventional audio-based approaches}
The spontaneous speech produced by HC, MCI and  AD speakers has different acoustic characteristics 
\cite{szatloczki2015speaking,Syed2020}.
Features that differentiate these groups can be captured by a fixed-dimensional representation. 
The present study employs two types of audio-based baseline systems to extract the representations. 
The first one is based on paralinguistic features extracted using the OpenSMILE toolkit \cite{eyben2010opensmile}.
Frequency, energy, and spectral aspects of the signal, called the low-level descriptors (LLD), are computed at frame-level. Statistic functionals are applied to the LLDs to derive a large set of parameters for each utterance. The paralinguistic features of IS10 \cite{schuller10b_interspeech}, ComParE2016 \cite{Schuller+2016} and eGeMAPS \cite{eyben2015geneva} are used in the baseline system.

The second type of the baseline system utilizes the speaker verification (SV) approach via deep neural network (DNN).
Speaker representations that carry discriminative information of speaker groups (AD/MCI/HC) are extracted from the DNN. 
We employ the state-of-the-art ECAPA-TDNN \cite{desplanques2020ecapa} as the DNN architecture. The model is pre-trained on Voxceleb2 \cite{chung2018voxceleb2} and CN-Celeb \cite{fan2020cn}.
The speaker representations are classified as AD/MCI/HC by a fully-connected layer.


\subsubsection{Transcript-based approaches}
\label{sec:trans-approach}
Previous studies \cite{DBLP:conf/interspeech/LuzHFFM20,pappagari20_interspeech} showed that transcript-based features extracted from speech transcripts were able to outperform the conventional audio-based features in the AD recognition task.  Given speech transcripts as input, BERT and BERT-like models fine-tuned by AD recognition task were successfully applied to detect AD for English-speaking subjects  \cite{pappagari20_interspeech,yuan2020disfluencies,DBLP:conf/interspeech/BalagopalanERN20}. This motivates the use of such pre-trained language models to perform AD recognition for our  Chinese-speaking subjects. 
In the present study, speech transcripts in the AD dataset are manually annotated following the CHAT protocol \cite{macwhinney2000childes}. Note that the speech sample AD\_M\_230706\_001 is excluded due to the difficulty in annotation. 
BERT and RoBERTa based models that support Chinese are adopted for AD recognition, including \emph{bert-base-chinese} developed by Google \cite{DBLP:conf/naacl/DevlinCLT19} and \emph{chinese-bert-wwm-ext}, \emph{chinese-RoBERTa-wwm-ext}, \emph{chinese-RoBERTa-wwm-ext-large} developed by HFL \cite{chinese-bert-wwm}. The last three models applied Chinese word segmentation and whole word masking (wwm) techniques to the vanilla BERT/RoBERTa model, achieving significant performance improvement in numerous Chinese NLP tasks \cite{chinese-bert-wwm}. To lift the dependence on human-prepared transcripts, automated-transcripts produced by ASR systems are also used to fine-tune the pre-trained language models for AD recognition. ASR systems trained with two large-scale Chinese corpora , namely AISHELL1 \cite{bu2017aishell} and AISHELL2 \cite{du2018aishell}, are employed to recognize the speech from AD dataset.


\subsection{Proposed approach: transferring pre-trained ASR models for AD recognition}

\begin{figure}[H]
\centering
\includegraphics[width=\linewidth]{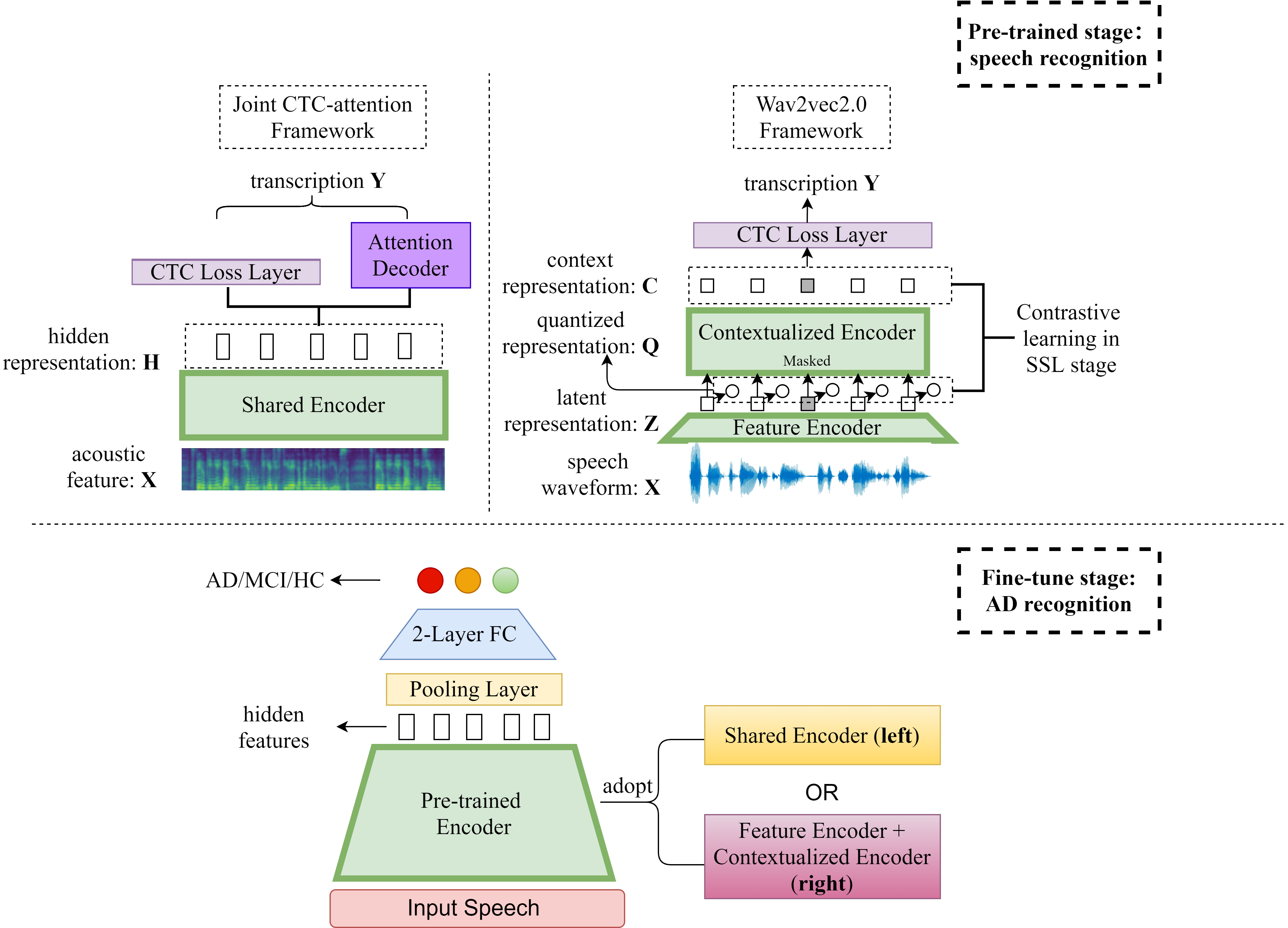}
\caption{The overall diagram of our proposed approach: transfer ASR models for AD recognition. 
\label{proposed_flowchart}}
\end{figure}

Despite the fact that transcripts generated from an ASR model are in general inaccurate for impaired speech, the ASR model is still expected to be useful to extract linguistic information for AD detection. As shown on the top of Figure \ref{proposed_flowchart}, two state-of-the-art E2E ASR models are selected, namely, the joint CTC-attention model and the wav2vec2.0 based model. Both models are pre-trained for the speech recognition task. Then only the encoder parts (in green color) are kept and appended with a pooling layer and several fully-connected layers. Finally, the integrated model is fine-tuned for the downstream AD recognition task.

\subsubsection{Leveraging the joint CTC-attention ASR model}
\label{subsec:ctc-asr}
The joint CTC-attention model, as a typical supervisedly trained ASR model, is adopted. It comprises three modules, namely a shared encoder,  an attention decoder and a connectionist temporal classification (CTC) loss layer.  As shown in the top-left of Figure \ref{proposed_flowchart}, an input sequence of acoustic features $\mathbf{X}$ is encoded into a hidden representation $\mathbf{H}$ by the shared encoder.  The hidden sequence is then processed in parallel by the attention decoder and the CTC loss layer  to generate the final transcripts $\mathbf{Y}$.  The three modules are jointly optimized for speech recognition. 

\begin{figure}[ht]
\includegraphics[width=0.75\linewidth]{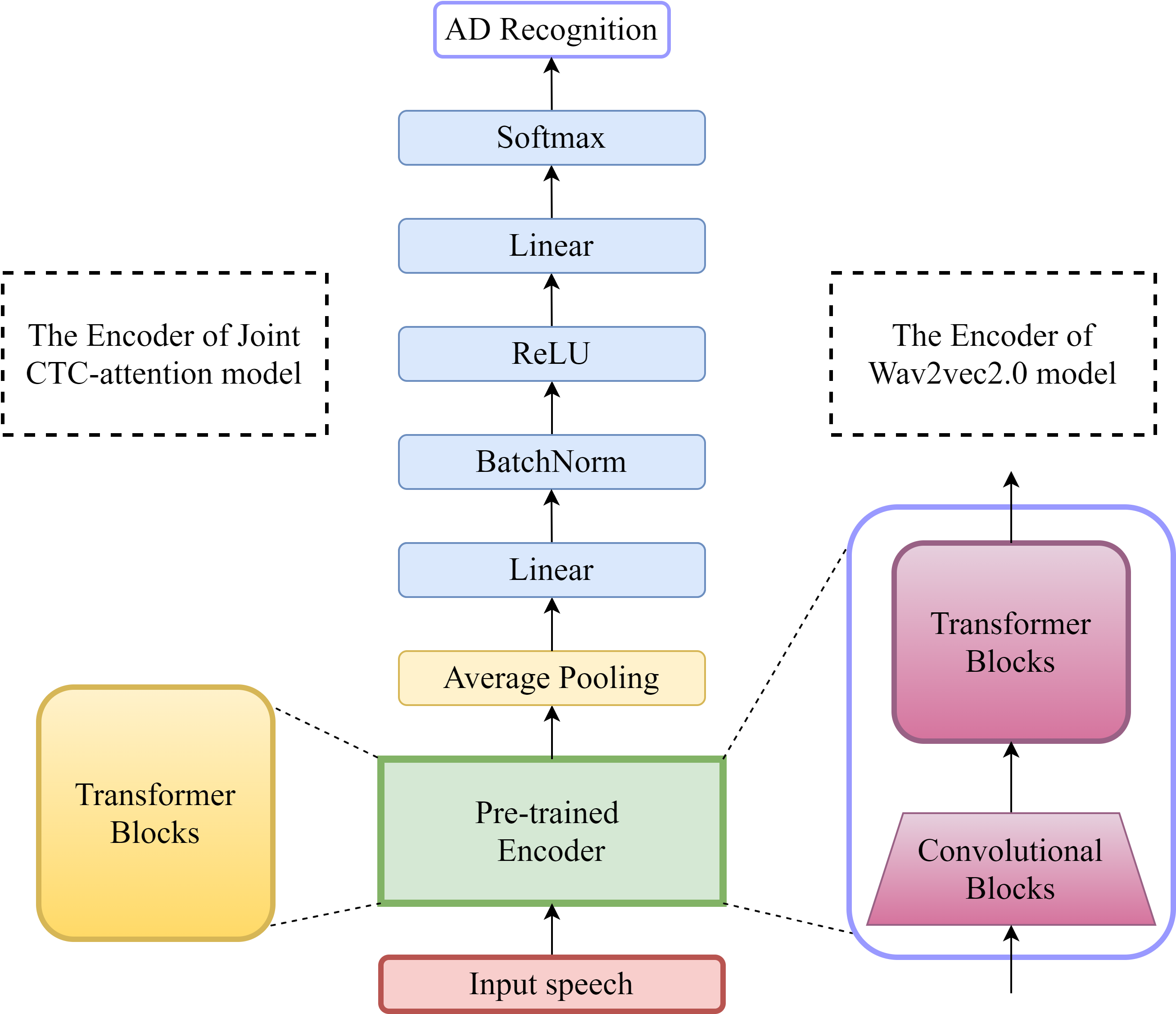}
\centering
\caption{The integrated AD recognition model that leverages the pre-trained encoder of ASR models. The pre-trained encoder module can come from the joint CTC-attention model (left) or wav2vec2.0 model (right).\label{fig:detailed_model}}
\end{figure}

In the fine-tune stage, only its shared encoder module (in green color) is kept for the downstream AD recognition task. An average pooling layer followed by a two-layer feed-forward neural network is appended to the encoder. Figure \ref{fig:detailed_model} illustrates the resulting integrated model in detail .

\subsubsection{Leveraging the wav2vec2.0 based ASR model}

The wav2vec2.0 model is a recent breakthrough of self-supervised speech representation learning (SSL) \cite{baevski2020wav2vec}. It has been shown to discover some discrete underlying speech units directly from raw speech data without the need of transcripts in the training process. This makes it possible to develop a state-of-the-art ASR model with  limited amount of labeled speech data. 

The model wav2vec2.0 mainly consists of three components, namely feature encoder, contextualized encoder and quantization module. In the forward pass, as illustrated in the top-right of Figure \ref{proposed_flowchart}, the feature encoder takes the raw speech waveform $\mathbf{X}$ as input and output a sequence of compact latent representations $\mathbf{Z}$. Then they are fed into a contextualized encoder to generate the final context representation $\mathbf{C}$ which aims to capture the long range dependencies over the entire sequence. 
With a quantization module, the latent representations $\mathbf{Z}$ are discretized into a finite set of quantized representation $\mathbf{Q}$ to represent the learning targets in the SSL pre-training stage. The object of SSL is to distinguish the correct quantized representation $q_t$ from distractors at each masked time step $t$ via contrastive learning.

Since our AD dataset is in Chinese, the cross-lingual speech representation (XLSR) version of wav2vec2.0  \cite{conneau2020unsupervised} is used in this study. After self-supervised training, a randomly initialized projection layer is added on the top of wav2vec2.0 for supervised ASR training. The projection layer is used to map the context representation $\mathbf{C}$ to the vocabulary of the target language in transcription $\mathbf{Y}$. The whole network is jointly trained with the CTC loss. In this case, the wav2vec2.0 can be regarded as a kind of acoustic model with pre-encoded linguistic information. 
Only its feature encoder (convolutional blocks) and contextualized encoder (transformer blocks) are kept for fine-tuning on the AD recognition task. Similar to the approach  of integrating the joint CTC-attention ASR model, an average pooling layer and a two-layer fully-connected network are  stacked to the wav2vec2.0 for AD recognition. 

\section{Experimental Setup}
\label{sec:exp-setup}
\subsection{Implementation configurations} 
\label{subsec:config}
Table \ref{tab:exp_config} summarizes the configurations for all baselines and our proposed approaches. 
A data augmentation method, i.e., splitting audio samples or speech transcripts into small segments, is applicable for both training and evaluation. If applied in the test stage, the sample-level prediction can be derived from majority voting on the associated segment-level predictions. We found that this augmentation method does not benefit all the experimental approaches. Table \ref{tab:exp_config} lists the best augmentation setup for each individual approach. More implementation details are described as follows:
\begin{enumerate} 
    \item \textbf{Conventional audio-based approach:} 
    For the first baseline system, the paralinguistic features are extracted using the OpenSMILE toolkit \cite{eyben2010opensmile}. A support vector machine (SVM) with linear kernel is trained on each type of feature. Mean and variance normalisation is applied to the paralinguistic features. The complexity parameter of the SVM is set to $1$. 
    
    For the second baseline system, ECAPA-TDNN with $1024$ channels is employed. The model is pre-trained on Voxceleb2 \cite{chung2018voxceleb2} and CN-Celeb \cite{fan2020cn} in sequence. In each dataset, the model is trained for $20$ epochs with Adam optimizer and batch size $=128$. The initial learning rate is $1e-3$ and decays by $10$ times every $8$ epochs. The log mel-filterbank (MBank) is taken as the model input, which is transformed from the input speech with window size $=25ms$ and window shift $=10ms$. In the fine-tune stage for AD recognition, the speaker classification layer (the last fully-connected layer) is replaced by a fully-connected layer with three-class output, and the whole model is trained for $20$ epochs. During training, a 6-second duration segment is randomly cropped from each MBank as model input. In the evaluation, one sample is segmented into 6-second duration with 5-second overlap and the averaged segments' output is utilized as the sample-level output.
    \item \textbf{Transcript-based approach:} All pre-trained language models are implemented with the HuggingFace Transformers Library \cite{wolf-etal-2020-transformers}. A three-dimensional linear layer with softmax activation function is added to pre-trained Chinese BERT/RoBERTa models to perform AD recognition. Cross-entropy loss is adopted as the loss function. The following hyperparameters are selected empirically to maximize the classification accuracy of development set: batch size = $8$, learning rate = $1e-5$, optimizer = Adam, max input length = $512$, epochs are set in the range of [$20$, $30$] using the early stopping strategy for different models. Our preliminary experimental results suggest that the \emph{chinese-bert-wwm-ext} and \emph{chinese-RoBERTa-wwm-ext} fine-tuned with full-length transcripts of speech samples perform better than other models, such that the results given by these two models will be reported in the following sections. As mentioned in section \ref{sec:trans-approach}, the input to pre-trained language models can be either manual transcripts or ASR output. For the generation of automated-transcripts, the joint CTC-attention ASR model described in section \ref{subsec:ctc-asr} is trained with AISHELL1 and AISHELL2, achieving a character error rate (CER) of $79.9\%$ and $68.3\%$ respectively. 
    \item \textbf{The proposed approach by transferring pre-trained ASR models:} The encoders from ASR models are fine-tuned with cross-entropy loss for the AD recognition task. The representation extracted from the last layer of the ASR encoder is used by default. For the joint CTC-attention ASR, an AISHELL1 pre-trained model \footnote{https://zenodo.org/record/4604023\#.YU7ZSY4zaUk}is adopted. The following hyperparameters are used in the fine-tune stage: batch size = $32$, optimizer = Adam, learning rate = $1.5e-3$, max-epoch = $20$. The training audio samples are segmented into 3-second long with a hop size of $1$ second. For the wav2vec2.0 based ASR, a pre-trained model \footnote{https://huggingface.co/jonatasgrosman/wav2vec2-large-xlsr-53-chinese-zh-cn} from the Huggingface Transformer Library is adopted. It fine-tunes the raw SSL model \footnote{https://huggingface.co/facebook/wav2vec2-large-xlsr-53} ``facebook/wav2vec2-large-xlsr-53'' on Chinese mainly using the Common Voice corpus \cite{ardila2019common}. In the AD fine-tune stage, the hyperparameters are set as: batch size = $32$, optimizer = AdamW, initial learning rate = $1e-3$ with a linear decay scheduler, max-epoch = $50$ with the early stopping strategy depending on the evaluation performance on  the \textit{dev} set. Data augmentation is applied in both training and test stages. The audio samples are randomly cropped into $10$-second segments for training, while different sizes of segment and hop length are used for evaluation (see details in Table \ref{figures_table}).  
\end{enumerate}

 \begin{specialtable}[h]
\caption{Summary of the best model configurations for all approaches. Here ``Use data augmentation'' indicates whether the model takes the segments (Yes) or full-length
samples (No) as inputs during the training and test stages. \label{tab:exp_config}}
\scalebox{0.8}{
\begin{tabular}{c|c|c|cc}
\toprule
\multirow{2}{*}{Category}                                                                                           & \multicolumn{2}{c|}{\multirow{2}{*}{Model Configuration}}                                                                   & \multicolumn{2}{c}{Use data  augmentation} \\ \cline{4-5} 
                                                                                                                    & \multicolumn{2}{c|}{}                                                                                                & \multicolumn{1}{c|}{train}    & test    \\ \hline
\multirow{4}{*}{\begin{tabular}[c]{@{}c@{}} Conventional \\ audio-based\end{tabular}}                                 & \multirow{3}{*}{\begin{tabular}[c]{@{}c@{}} OpenSMILE features\end{tabular}} & IS10 & \multicolumn{1}{c|}{Yes}      & Yes     \\ \cline{3-5} 
                                                                                                                    &                                                                                             & ComParE2016            & \multicolumn{1}{c|}{Yes}      & No      \\ \cline{3-5} 
                                                                                                                    &                                                                                             & eGeMAPS                & \multicolumn{1}{c|}{Yes}      & No      \\ \cline{2-5} 
                                                                                                                    & \multicolumn{2}{c|}{ECAPA-TDNN (Voxceleb2 and CN-Celeb)}                                                             & \multicolumn{1}{c|}{Yes}      & Yes     \\ \hline
\multirow{3}{*}{Transcript-based}                                                                                   & \multirow{3}{*}{Chinese BERT/RoBERTa}                                                    & Manual transcripts                & \multicolumn{1}{c|}{No}       & No      \\ \cline{3-5} 
                                                                                                                    &                                                                                             & ASR output (AISHELL1)          & \multicolumn{1}{c|}{No}       & No      \\ \cline{3-5} 
                                                                                                                    &                                                                                             & ASR output (AISHELL2)          & \multicolumn{1}{c|}{No}       & No      \\ \hline
\multirow{3}{*}{\textbf{\begin{tabular}[c]{@{}c@{}}\\ Proposed:\\transfer pre-trained\\ ASR model\end{tabular}}} & \multicolumn{2}{c|}{Joint CTC-attention (AISHELL1)}                                                                  & \multicolumn{1}{c|}{Yes}      & No      \\ \cline{2-5} 
                                                                                                                    & \multirow{2}{*}{\begin{tabular}[c]{@{}c@{}}Wav2vec2.0\\ (XLSR-zh-CN)\end{tabular}}          & \begin{tabular}[c]{@{}c@{}}Encoder for classification:\\ last layer \end{tabular} & \multicolumn{1}{c|}{Yes}      & Yes     \\ \cline{3-5} 
                                                                                                                    &                                                                                             & \begin{tabular}[c]{@{}c@{}}Encoder for classification:\\ concatenation of last 3 layers \end{tabular}            & \multicolumn{1}{c|}{Yes}      & Yes     \\ \bottomrule
\end{tabular}}
\end{specialtable}
 
\subsection{Evaluation metrics}

To measure the model performance of AD recognition, four commonly used evaluation metrics including  accuracy, macro-averaged precision, recall and F1 score are adopted. The perfect performance is attained when all metrics are equal to $1$.

\section{Experimental Results and Discussion}
\label{sec:exp-result}

\subsection{Conventional audio-based approach vs. Transcript-based approach}
As shown in the first block of Table \ref{figures_table}, four kinds of conventional audio-based models are listed. The top three rows represent OpenSMILE series with different feature types. It was shown that the IS10 features achieve the best performance on the \textit{test} set among all  OpenSMILE features. The ECAPA-TDNN based model obtains around $70.5\%$ accuracy on the \textit{test} set, which slightly outperforms the IS10 features. In the middle block of Table \ref{figures_table}, two different BERT-like models, i.e., ``BERT\_manual'' and ``RoBERTa\_manual'', which operate on the manual speech transcripts are shown. Here, ``BERT'' and ``RoBERTa'' specifically represent the pre-trained models \emph{chinese-bert-wwm-ext} and  \emph{chinese-RoBERTa-wwm-ext}, respectively. Compared to the BERT model, the RoBERTa shows a significant performance gain based on manual transcripts, which suggests the pre-trained Chinese RoBERTa model is more appropriate to be fine-tuned on  the AD recognition task than the BERT model.  
The best-performing audio-based model (ECAPA-TDNN) shows inferior to the ``RoBERTa\_manual''. This implies that the prominent characteristics of AD can be better captured by the transcript-based approaches than the conventional audio-based approaches. 

\begin{specialtable}[t]
\caption{The overall performance of AD recognition using conventional audio-based models, transcript-based models and our proposed models. For each type of evaluation metrics, the mean value and standard deviation value computed across v1/v2/v3 splits are reported for the \emph{dev} set and \emph{test} set respectively. 
The suffixes that follow ``Wav2vec2.0'' indicate different settings of segment size and hop length during the test stage, e.g., ``3-2'' means 3s-segment with 2s-hop is adopted. The model ``Wav2vec2.0-last3'' follows the ``3-2'' setting, in which the suffix ``last3'' indicates that the last three layers of the wav2vec2.0  encoder are concatenated for further classification, instead of the last layer. }

\label{figures_table}
\scalebox{0.74}{
\begin{tabular}{c|c|c|c|c}
\toprule
\multirow{2}{*}{Model}           & \multicolumn{4}{c}{Evaluation Metrics (\%)}                                                                                              \\ \cline{2-5} 
    &Accuracy (dev/test)                        & Precision (dev/test)                       & Recall (dev/test)                           & F1 (dev/test)                              \\ \hline
IS10  & 84.9$\pm$7.0 / 69.0$\pm$1.1  & 84.6$\pm$7.6 / 71.1$\pm$0.1  & 84.6$\pm$7.1 / 67.7$\pm$1.2  & 84.3$\pm$7.2 / 68.3$\pm$1.2  \\
                                                                                                                  ComParE2016 & 83.9$\pm$7.0 / 61.2$\pm$1.1 & 83.9$\pm$8.1 /  62.5$\pm$1.3  & 83.0$\pm$7.4 / 58.1$\pm$1.3  & 83.0$\pm$7.5 / 56.6$\pm$2.2  \\
                                                                                                                  eGeMAPS & 83.0$\pm$5.6 / 58.1$\pm$7.0  & 82.8$\pm$6.0 / 63.7$\pm$9.3 & 82.5$\pm$5.6 / 60.2$\pm$8.0  & 82.5$\pm$5.8 / 60.3$\pm$8.7  \\
                                                                                                                  \textbf{ECAPA-TDNN}                             & 82.0$\pm$8.6 / \textbf{70.5}$\pm$2.9  & 82.9$\pm$8.0 / 72.8$\pm$4.1  & 79.8$\pm$10.0 / 69.0$\pm$2.3 & 79.9$\pm$9.8 / 68.4$\pm$2.5  \\ \hline
BERT\_manual     & 82.5$\pm$7.4 / 69.8$\pm$0.0  & 82.1$\pm$8.2 / 70.3$\pm$0.3  & 81.4$\pm$8.0 / 67.8$\pm$0.2  & 81.3$\pm$8.2 / 67.8$\pm$0.5  \\
                                                                                                                  \textbf{RoBERTa\_manual}  & 86.3$\pm$6.2 / \textbf{75.2}$\pm$2.9  & 86.0$\pm$6.8 /  77.2$\pm$2.9  & 84.9$\pm$8.3 / 74.0$\pm$3.1  & 84.7$\pm$8.2 / 74.7$\pm$3.1  \\
                                                                                                                  RoBERTa\_aishell1                         & 59.9$\pm$2.8 / 50.4$\pm$1.1  & 59.5$\pm$2.3 / 48.6$\pm$3.0  & 58.0$\pm$3.4 / 47.1$\pm$0.9  & 57.9$\pm$3.1 / 43.0$\pm$0.7   \\
                                                                                                                  RoBERTa\_aishell2                        & 78.1$\pm$10.5 / 53.5$\pm$1.9 & 78.5$\pm$10.4 / 51.3$\pm$2.1 & 78.6$\pm$10.7 / 51.6$\pm$2.0 & 78.1$\pm$10.5 / 51.1$\pm$2.1 \\ \hline
Joint CTC-attention                              & 87.9$\pm$5.7 / 79.1$\pm$1.9  & 88.0$\pm$5.8 / 78.4$\pm$1.9  & 87.9$\pm$5.5 / 77.3$\pm$2.1  & 87.6$\pm$5.7 / 77.3$\pm$2.3  \\
                                                                                                                  \textbf{Wav2vec2.0\_3-2}                        & 87.9$\pm$5.7 / \textbf{79.8}$\pm$4.0  & 87.9$\pm$5.7 / 80.7$\pm$5.0  & 88.4$\pm$5.1 / 79.7$\pm$4.2 & 87.7$\pm$5.6 / 79.6$\pm$4.3  \\
                                                                                                                  Wav2vec2.0\_6-5                        & 86.6$\pm$3.3 / 79.1$\pm$3.8  & 87.7$\pm$3.7 / 80.6$\pm$4.2  & 87.5$\pm$2.4 / 79.2$\pm$3.9  & 86.6$\pm$3.2 / 79.1$\pm$3.9  \\
                                                                                                                  Wav2vec2.0\_10-5                       & 87.5$\pm$5.5 / 79.1$\pm$3.3  & 87.6$\pm$5.4 / 79.8$\pm$4.4  & 88.1$\pm$4.8 / 78.8$\pm$3.7  & 87.4$\pm$5.4 / 78.7$\pm$3.7  \\
                                                                                                                  Wav2vec2.0\_15-5                       & 87.5$\pm$5.5 / 77.5$\pm$2.2  & 88.1$\pm$5.7 / 78.3$\pm$3.6  & 88.1$\pm$4.8 / 77.3$\pm$2.7  & 87.4$\pm$5.4 / 77.2$\pm$2.7  \\
                                                                                                                 Wav2vec2.0-last3                       & 88.3$\pm$3.4 / 77.5$\pm$7.7  & 88.3$\pm$4.0 / 78.4$\pm$7.2  & 88.6$\pm$3.5 / 76.6$\pm$8.1  & 87.8$\pm$3.6 / 77.1$\pm$7.9  \\ \bottomrule
\end{tabular}}
\end{specialtable}

\subsection{Manual-transcript vs. ASR-transcript based pre-trained language model}
To make a fully automatic pipeline of transcript-based AD recognition, it is straightforward to replace the manual annotation with an ASR system. In this experiment, two general-purpose Chinese ASR systems trained with AISHELL1 and AISHELL2 are adopted (see section \ref{subsec:config}). The best RoBERTa model is used to process the transcripts generated by the ASR systems, namely ``RoBERTa\_aishell1'' and ``RoBERTa\_aishell2''. As illustrated in the middle block of Table \ref{figures_table}, the ``RoBERTa\_aishell2'' performs better than the ``RoBERTa\_aishell1'', achieving an averaged accuracy of $53.5\%$. This may be due to the lower CER attained by the AISHELL2-trained ASR. We also notice that their performance is much worse than their counterparts fine-tuned with manual transcripts. Our AD dataset contains spontaneous speech that comes from subjects with various accents, dialects and voice quality, which would severely affect the quality of automated-transcripts and thus leading to worse AD recognition results. This calls for a domain-specific ASR to deal with such high speaker variability. 



\subsection{Proposed approach vs. Conventional approaches}
In the bottom of Table \ref{figures_table}, the performance of our proposed approach by transferring pre-trained ASR models is reported. The joint CTC-attention based model and wav2vec2.0 based models with different settings of segment size and hop length are compared. It is worthy noted that all proposed models perform better than the golden transcript-based model  ``RoBERTa\_manual''. The highest accuracy of $79.8\%$ is obtained with the ``Wav2vec2.0\_3-2'' model, outperforming the  ``RoBERTa\_manual'' by an absolute gain of $4.6\%$. Such notable performance improvement may benefit from the joint fine-tuning  between front-end ASR and back-end AD recognition. Specifically, our proposed approach discards the explicit speech transcription procedure, and only linguistic information pre-encoded in the ASR is required, making the whole AD recognition process in an end-to-end manner. For the effect of different data augmentation settings on classification performance, it is found that aggregating predictions of shorter segments will give better performance than those of longer segments, e.g., ``Wav2vec2.0\_3-2'' surpasses ``Wav2vec2.0\_15-5'' by an absolute accuracy gain of $2.3\%$. However, the training stability is in an opposite way, i.e., ``Wav2vec2.0\_3-2'' has a larger standard deviation (std.) value than others. The model ``Wav2vec2.0-last3''  also shows such instability with a std. of $7.7\%$ in terms of accuracy. With the same data augmentation as ``Wav2vec2.0\_3-2'',  it gives worse performance on the \textit{test} set but shows the contrary on the \textit{dev} set. We suspect the tendency of overfitting has occurred by concatenating the last three-layer  representations.

\begin{figure}[H]
\centering
\includegraphics[width=0.7\linewidth]{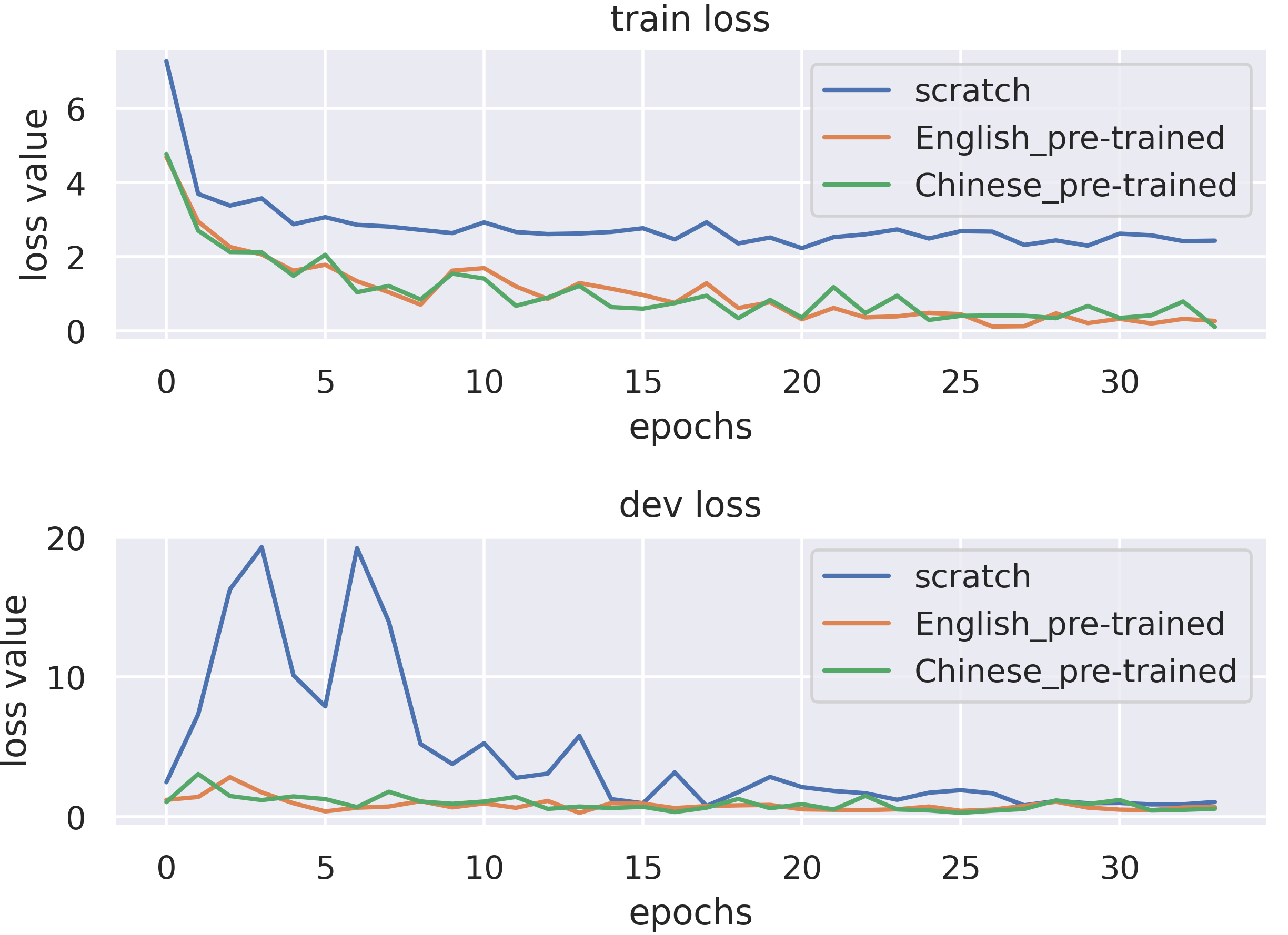}
\caption{The comparison of three models in terms of loss curve. The above are loss curves for training and the below are loss curves for validation. Note that all the loss curves  are recorded from experiments based on the v1 split.\label{fig:loss_comparison}}
\end{figure}

\begin{figure}[h]
\captionsetup[subfigure]{justification=centering}
     \centering
     \begin{subfigure}[b]{0.23\textwidth}
         \centering
         \includegraphics[width=0.85\textwidth]{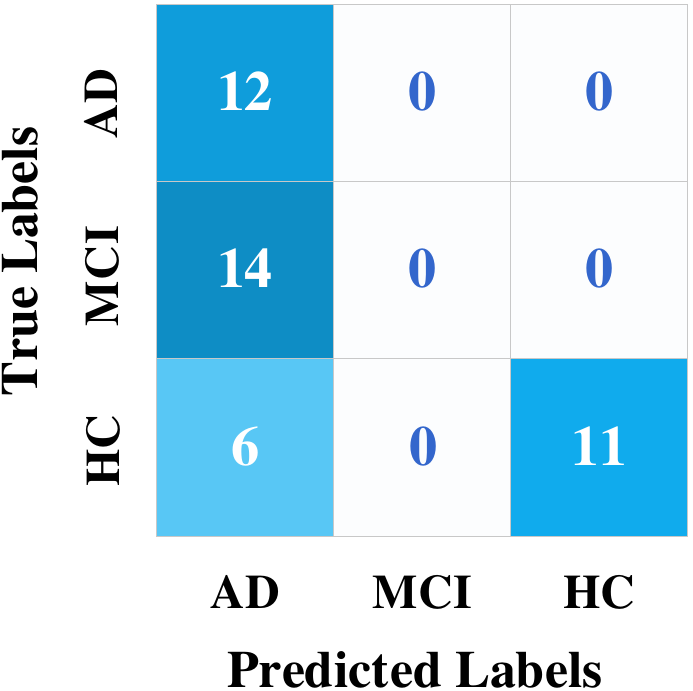}
         \caption{scratch}
         \label{fig:scratch}
     \end{subfigure}
     \hfill
     \begin{subfigure}[b]{0.2\textwidth}
         \centering
         \includegraphics[width=0.85\textwidth]{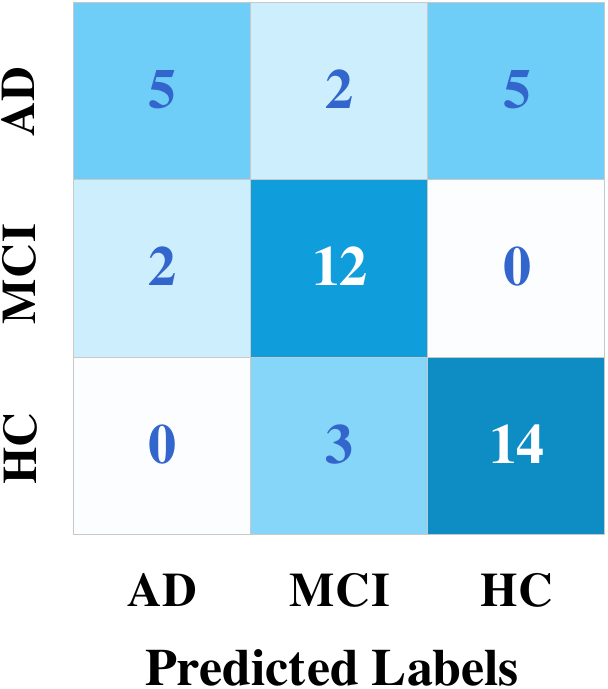}
         \caption{English\_pre-trained}
         \label{fig:pretrain-English}
     \end{subfigure}
     \hfill
     \begin{subfigure}[b]{0.2\textwidth}
         \centering
         \includegraphics[width=0.85\textwidth]{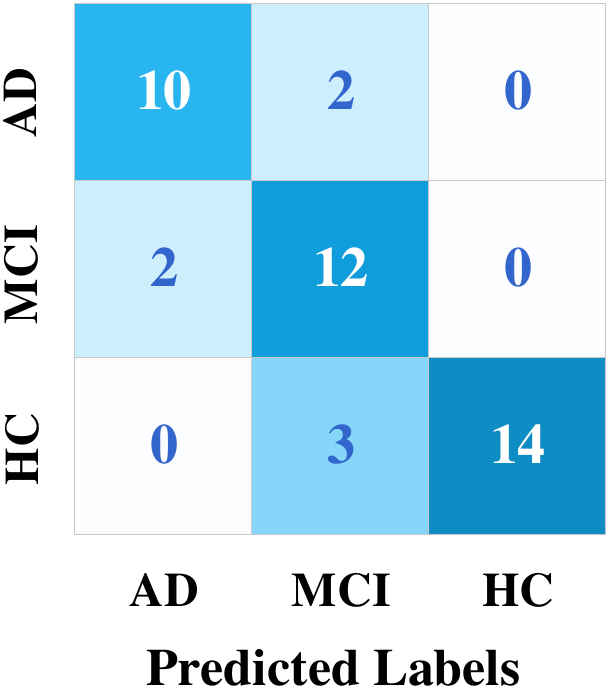}
         \caption{Chinese\_pre-trained}
         \label{fig:pretrain-Chinese}
     \end{subfigure}
        \caption{Comparing confusion matrices of AD recognition results given by three models. }
        \label{fig:three graphs}
\end{figure}

In order to better understand the advantages of using a pre-trained Chinese ASR
model, we experiment with a simple ablation study. 
With the same wav2vec2.0 based structure, three models are compared: (1) wav2vec2.0 trained from scratch (``scratch''); (2) wav2vec2.0 fine-tuned from a pre-trained English ASR; 
(3) wav2vec2.0 fine-tuned from a pre-trained Chinese ASR (``Chinese\_pre-trained''). All models target at the AD recognition task. 
In Figure \ref{fig:loss_comparison}, we can observe that the model trained from scratch is difficult to converge compared with the other two models. From the loss curves, the  ``English\_pre-trained'' seems to perform similarly with the ``Chinese\_pre-trained''. 
However, it should be noted that the ability of generalization is different for these two models.
To show it clearly, the confusion matrices of the above three models are computed on our unseen \textit{test} set, as depicted in Figure \ref{fig:three graphs}. It can be seen that the ``English\_pre-trained'' model cannot differentiate the AD subjects well. 
It recognizes $5$ out of $12$ AD subjects as healthy people. 

\subsection{\textbf{Results submitted to the AD Recognition Challenge}}
Two competition tracks are set in the AD Recognition Challenge. One aims at AD recognition using long speech audio ($119$ samples, $60$-second per sample) and the other using short speech audio ($1153$ samples, $6$-second per sample). Each participant team has three submission chances for each track. Three proposed models including ``Joint CTC-attention'', ``Wav2vec2.0\_3-2'' and ``Wav2vec2.0-last3''  are selected to give predictions respectively. 
Table \ref{tab:submission} compares the performance of our proposed models and the official baseline models. 
All models perform better in the long-audio competition track than the  short-audio track. This suggests that automatic AD recognition is better carried out with long speech samples. 
In terms of accuracy, the model ``Wav2vec2.0\_3-2'' achieves the best performance $\mathbf{83.2}\%$ in the track of long audio and the model ``Joint CTC-attention'' obtains the highest accuracy of $\mathbf{78.0}\%$ in the track of short audio. It can be seen that our proposed models beat all baselines  \footnote{https://github.com/THUsatlab/AD2021} using OpenSMILE features (long-audio track) and convolutional neural network (short-audio track) provided by the organizer.  

\begin{figure}[h]
\captionsetup[subfigure]{justification=centering}
     \centering
     \begin{subfigure}[b]{0.23\textwidth}
         \centering
         \includegraphics[width=0.85\textwidth]{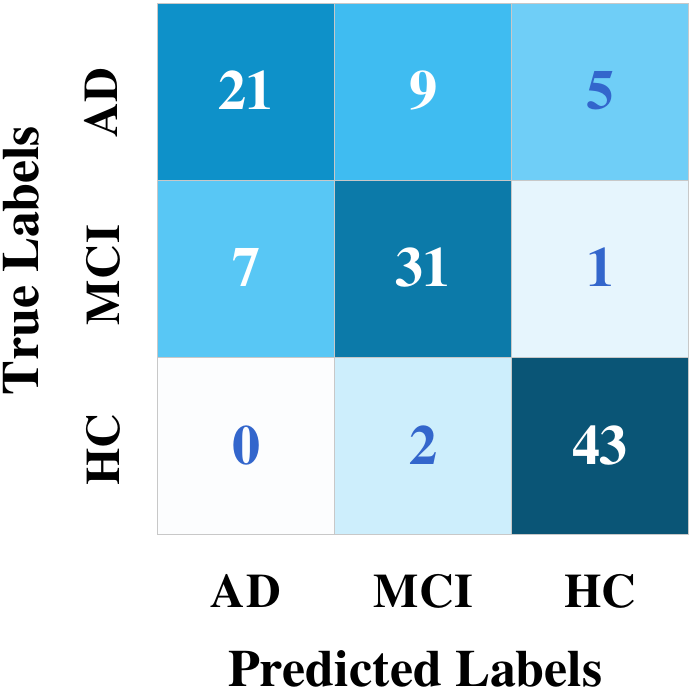}
         \caption{Joint CTC-attention}
     \end{subfigure}
     \hfill
     \begin{subfigure}[b]{0.2\textwidth}
         \centering
         \includegraphics[width=0.85\textwidth]{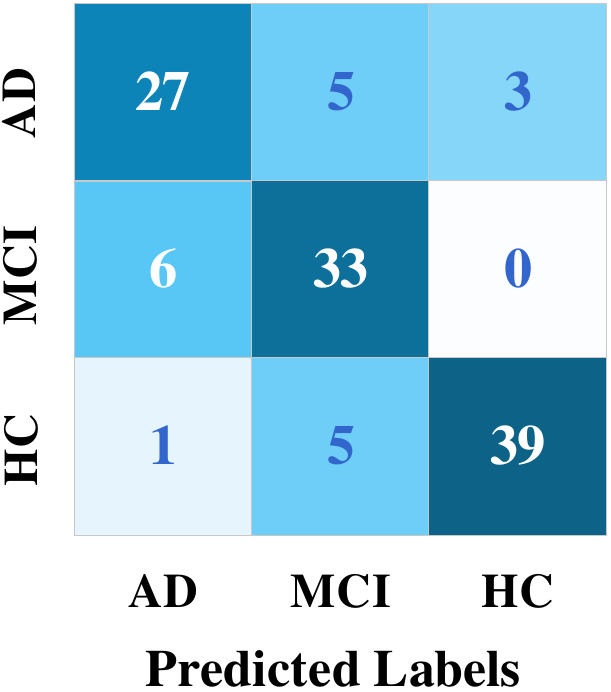}
         \caption{Wav2vec2.0\_3-2}
     \end{subfigure}
     \hfill
     \begin{subfigure}[b]{0.2\textwidth}
         \centering
         \includegraphics[width=0.85\textwidth]{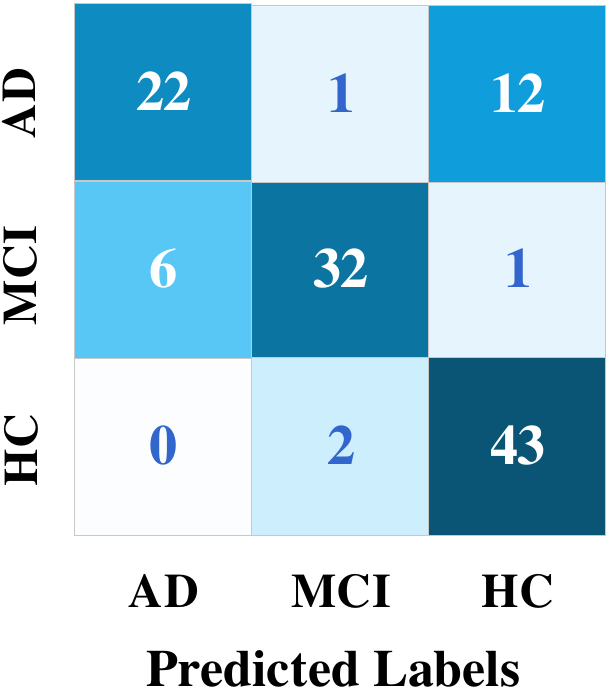}
         \caption{Wav2vec2.0-last3}
     \end{subfigure}
        \caption{The confusion matrices from three submitted models in the track of long audio. }
        \label{fig:cm_long}
\end{figure}

\begin{figure}[h]
\captionsetup[subfigure]{justification=centering}
     \centering
     \begin{subfigure}[b]{0.23\textwidth}
         \centering
         \includegraphics[width=0.85\textwidth]{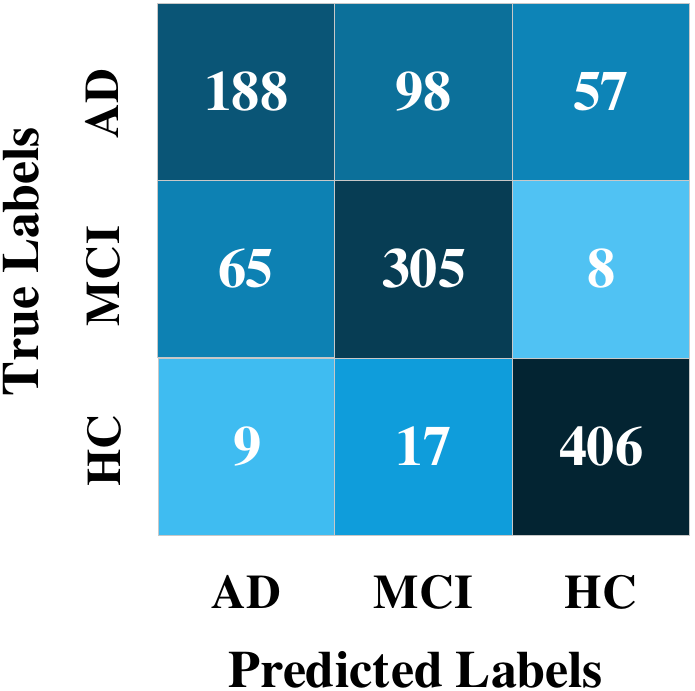}
         \caption{Joint CTC-attention}
     \end{subfigure}
     \hfill
     \begin{subfigure}[b]{0.2\textwidth}
         \centering
         \includegraphics[width=0.85\textwidth]{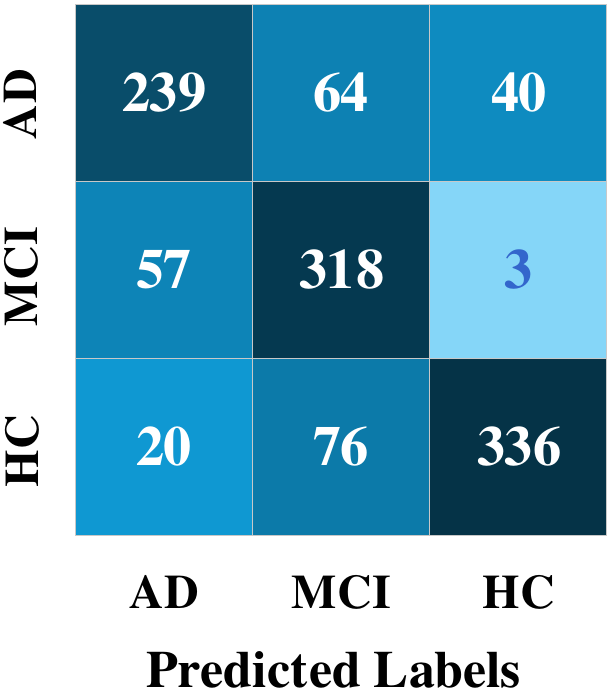}
         \caption{Wav2vec2.0\_3-2}
     \end{subfigure}
     \hfill
     \begin{subfigure}[b]{0.2\textwidth}
         \centering
         \includegraphics[width=0.85\textwidth]{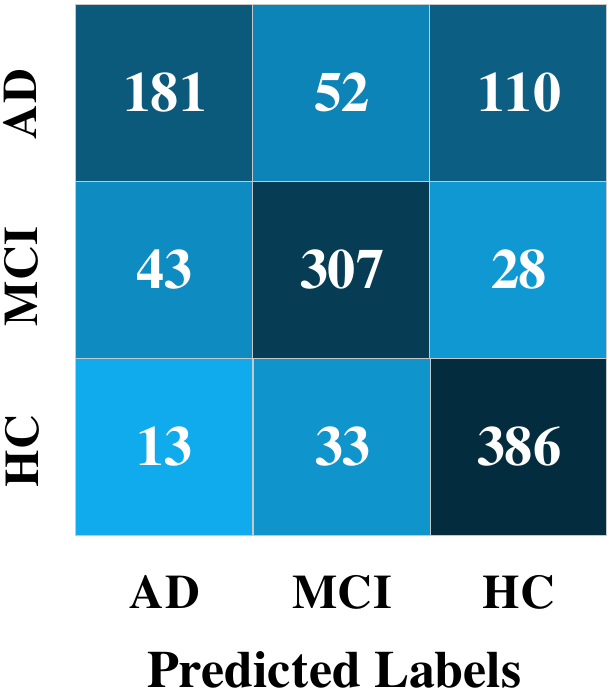}
         \caption{Wav2vec2.0-last3}
     \end{subfigure}
        \caption{The confusion matrices from three submitted models in the track of short audio. }
        \label{fig:cm_short}
\end{figure}

To investigate erroneous AD predictions made by the three models, 
their corresponding confusion matrices are computed and plotted. Figure \ref{fig:cm_long} shows the recognition results in the long-audio track and Figure \ref{fig:cm_short} gives the results in the short-audio track. For the results of long-audio track, we observe that all models tend to mis-classify AD as other two groups while the identification of HC seems to be the easiest task. The ``Wav2vec2.0\_3-2'' outperforms other models by correctly classifying more AD subjects. 
Recognition of AD using short audio is generally more difficult than using long audio due to less information embedded in short audio. The identification of AD is still the toughest task in the short-audio track. Among three models, the ``Joint CTC-attention'' model performs the best in distinguishing HC subjects from other subjects, surpassing other two models in the short-audio track.


\begin{specialtable}[h]
\centering
\caption{The performance of the official baseline and three submitted models on the official unseen test sets. ``l'' and ``s'' stand for long-audio track and short-audio track, respectively. \label{tab:submission}}
\scalebox{0.85}{
\begin{tabular}{c|cccc}
\toprule
\multirow{2}{*}{Model} & \multicolumn{4}{c}{Evaluation Metrics (\%)}                                                                                                           \\ \cline{2-5} 
                      & \multicolumn{1}{c|}{Accuracy (l/s)} & \multicolumn{1}{c|}{Precision (l/s)} & \multicolumn{1}{c|}{Recall (l/s)} & F1 (l/s) \\ \hline
Official baseline      & \multicolumn{1}{c|}{79.8 / 74.0}           & \multicolumn{1}{c|}{79.9 / 72.3}            & \multicolumn{1}{c|}{78.5 / 73.7}         & 78.6 / 71.8     \\ \hline
\textbf{Joint CTC-attention}    & \multicolumn{1}{c|}{79.8 / \textbf{78.0}}           & \multicolumn{1}{c|}{78.9 / 76.9}            & \multicolumn{1}{c|}{78.4 / 76.5}         & 78.2 / 76.2     \\ \hline
\textbf{Wav2vec2.0\_3-2}        & \multicolumn{1}{c|}{\textbf{83.2} / 77.5}           & \multicolumn{1}{c|}{83.0 / 77.9}            & \multicolumn{1}{c|}{82.8 / 77.2}         & 82.8 / 77.2     \\ \hline
Wav2vec2.0-last        & \multicolumn{1}{c|}{81.5 / 75.8}           & \multicolumn{1}{c|}{82.3 / 76.1}            & \multicolumn{1}{c|}{80.2 / 74.5}         & 80.5 / 74.3     \\ \bottomrule
\end{tabular}}
\end{specialtable}

\section{Conclusions}
\label{sec:conclusion}
This paper has presented a novel and light-weight framework to perform AD recognition through spontaneous speech, in which a pre-trained ASR model is tailored and adapted to the intended task. Two representative E2E ASR models, namely the supervised joint CTC-attention model and the self-supervised wav2vec2.0 based model are investigated. Compared with the conventional audio-based and transcript-based approaches, the proposed one integrates the ASR encoder and the back-end classifier into a single model for AD recognition. This makes it convenient to be jointly fine-tuned in an end-to-end fashion. 
Experimental results show that the linguistic information is more important than paralinguistic information for the recognition of AD. 
Our proposed approach can achieve superior performance than the best transcript-based model using Chinese RoBERTa, without the dependence on manual transcripts. It is also found that the language-specific linguistic information pre-encoded in the ASR model is essential to the model fine-tuning for the downstream AD recognition task. 

\vspace{6pt} 




\funding{This research is funded by the Fundamental Research Funds for the Central Universities $2021$RC$244$ and is partially supported by the Sustainable Research Fund of The Chinese University of Hong Kong.}

\end{paracol} 
\reftitle{References}

\bibliography{reference.bib}

\end{document}